\documentstyle[11pt]{article}

   \let\d=\delta
  \let\h=\eta

\let\O=\Omega   
 \let\G=\Gamma  

\def\0{\over } \def\1{\vec } \def\2{{1\over2}} \def\4{{1\over4}}
\def\5{\bar } 
\def\6{\partial }
\def\7#1{{#1}\llap{/}}
\def\8#1{{\textstyle{#1}}} \def\9#1{{\bf {#1}}}

\def\({\left(} \def\){\right)} \def\<{\langle } \def\>{\rangle }
\def\[{\left[} \def\]{\right]}  
 
\def\NPB#1{Nucl. Phys. {\bf B#1}}

\def\PRD#1{Phys. Rev. {\bf D#1}}
\def\PLB#1{Phys. Lett. {\bf B#1}}

\newcommand{\bea}{\begin{eqnarray}}
\newcommand{\beal}[1]{\begin{eqnarray}\label{#1}}
\newcommand{\eea}{\end{eqnarray}}
\newcommand{\be}{\begin{equation}}
\newcommand{\ee}{\end{equation}}
\newcommand{\bel}[1]{\begin{equation}\label{#1}}

\def\vew#1{\left\<{#1}\right\>}

\def\1{\hskip 0.8 true cm }
\def\'{\hskip 0.5 true cm }
\begin{document}
\begin{titlepage}
\begin{flushleft}
\tt TUW 97-17
\end{flushleft}
\vspace{1cm}

\begin{center}
{\Large\bf
Reply to ``Proof of the Gauge Independence\\ 
of the Conformal Anomaly of Bosonic String\\ 
in the Sense of Kraemmer and Rebhan''\\}
\vfill
{\large Ulrike Kraemmer and Anton Rebhan}
\bigskip\bigskip

{\it
Institut f\"ur Theoretische Physik,
        Technische Universit\"at Wien,\\
        Wiedner Hauptstr. 8--10,
        A-1040 Vienna, Austria}
\vfill
\today
\bigskip
\end{center}
{}\hrulefill {}\\
\noindent{\bf Abstract}

\bigskip
\noindent
In two recent preprints (hep-th/9710131 and 9710132), Abe and Nakanishi
have claimed that the proof of the gauge independence of the conformal
anomaly of the bosonic string as given by us in 1988
was wrong. A similar allegation has been made concerning
our proof of the gauge independence of the sum of the ghost
number and Lagrange multiplier anomalies in non-conformal gauges.
In this short note we refute their criticism by explaining
the simple logic of our
proofs and
emphasizing the points that have been missed by Abe and Nakanishi.
\\{\rule{0mm}{0mm}}\hrulefill \\
\vspace*{\fill}
\end{titlepage}

The standard approach to covariant quantization of string theory
chooses the conformal gauge, which in the critical dimension
permits to eliminate the world-sheet
metric as a dynamical variable\cite{P}.
The gauge independence of the conformal anomaly, which determines the
critical dimension, has been investigated first in Ref.~\cite{HM}, but
only for algebraic gauges. In Ref.~\cite{DD}, the world-sheet anomalies
were explicitly calculated for the harmonic gauge, which appeared to
be of particular
interest because of a vanishing ghost-number anomaly. The conformal
anomaly on the other hand remained unchanged as expected. In Ref.~\cite{RK}
the present authors have confirmed these results, but have shown that
the eliminated ghost-number anomaly has just been shifted to an analogous
number-current anomaly for the now dynamical world-sheet metric;
the sum of the ghost-number and the so-called Lagrange-multiplier
anomalies turned out to be gauge independent. The gauge independence
of the conformal anomaly was also verified in Refs.~\cite{JL,FLP,DMvN},
which considered the background-covariant harmonic gauge, and in 
Refs.~\cite{BB,RK}, where non-background-covariant de Donder gauges  
were used.

In Ref.~\cite{KR} we have reviewed and extended the explicit
calculations of the various world-sheet anomalies and have given a
simple proof of the observed gauge independences based on BRS symmetry.
In a more general framework, the gauge independence of anomalies was 
subsequently confirmed in Refs.~\cite{KPSS,WK}.

Some years later, a seemingly contradictory result was published by
Abe and Nakanishi\cite{AN0}, 
who claimed that in non-conformal gauges the conformal
anomaly was undetermined due to ambiguities in the definition of
the energy-momentum tensor.\footnote{The explicit result of Ref.~\cite{AN0}
is in fact erroneous. As Abe and Nakanishi had to admit recently \cite{AN2},
they had overlooked infrared divergences which restrict their alleged
ambiguity to a sign ambiguity.}
Although later Ref.~\cite{WM} has pointed out that
their ambiguity disappears if one refrains from restricting to a
flat background prematurely, in Refs.~\cite{AN1,AN2} Abe and Nakanishi
have recently questioned the previous works that showed the gauge
independence of world-sheet anomalies. In particular they claimed
that the explicit proof in Ref.~\cite{KR} was based on false assumptions
and therefore wrong. (Not trusting the more general arguments of 
Refs.~\cite{KPSS,WK}, they presented an alternative proof which
is however so restricted that they cannot even ``compare two gauges which
have different Feynman rules''.)

In Ref.~\cite{KR} we have considered diffeomorphism and Weyl
gauge conditions of the form
\be
F_i^{mn}[\hat g] h_{mn} =0,\quad \O^{mn}[\hat g] h_{mn} =0,\quad{i,m,n=0,1}
\ee
where $h_{mn}=g_{mn}-\hat g_{mn}$ with $\hat g$ being a classical
background field.
Imposing these gauge conditions by Lagrange multipliers $b^i$ and $b$,
respectively, one finds that $h_{mn}$ and $(b^i,b)$ fields have only
mixed propagators. As a consequence, the one-loop effective action
is exact and the gauge-fixed action can be
linearized with respect to the {\em quantum} metric field $h_{mn}$.
Correspondingly, the BRS symmetry can be linearized, which moreover
implies abelianization.

As is well known,
while anomalies arise only in the context of renormalization, they
are contained already in the regularized effective action $\G$, which
is a functional of the background metric field $\hat g_{mn}$. Local
contributions to the latter can be changed by renormalization,
but non-local ones can not. Using this freedom to restore any
diffeomorphism invariance that a particular regularization scheme
may have violated \cite{RKK}, the effective action is proportional
to $\int d^2x \sqrt{-\hat g}\hat R \hat\Box^{-1}\hat R$,
and its prefactor determines the conformal anomaly.
Since however everything is fixed by the non-local contributions, one
only needs to consider the regularized effective action in order
to establish the gauge independence of the conformal anomaly.

An infinitesimal variation $\d F$ of the gauge condition $F$ can be
shown to give
\bel{dG}
\d \G[\hat g]=\vew{Q_{\mathrm BRS} (\bar c^i \d F_i^{mn} h_{mn})}
\ee
where only the linearized BRS charge is needed as explained above.
For the linearized BRS transformations there cannot appear any
anomalies, hence (\ref{dG}) is zero and the conformal anomaly
is gauge-fixing independent.

In Ref.~\cite{AN1}, Abe and Nakanishi remark firstly that this argument
cannot be correct, for it would imply that the anomaly itself was zero,
not only its gauge variations, if $\d F$ was replaced by $F$.
They do not provide further explanations, so one can only guess that
they have in mind that the gauge sector of the gauge fixed action is
BRS exact and that therefore one could delete the $\d$'s in (\ref{dG}).
But the effective action is not the expectation value of the classical
action in the sense of (\ref{dG}). Nor could one integrate (\ref{dG})
starting from $F\equiv0$, because there the effective action would
no longer be well-defined.

More concretely, Abe and Nakanishi then point out that variations of the
background field and BRS transformations do not commute. This is true,
but contrary to what they claim, our proof does {\em not} depend on this.
In order to determine the proportionality factor in
\be\G[\hat g]\propto\int d^2x \sqrt{-\hat g}\hat R \hat\Box^{-1}\hat R
\ee
it is sufficient to extract its bilinear terms in an expansion
of $\hat g_{mn}=\h_{mn}+\hat h_{mn}$. No more and no less
is done in the explicit calculations in Refs.~\cite{DD,RK,KR}.

The very same strategy is followed in our treatment \cite{KR} 
of number-current anomalies.
These anomalies are contained in an effective action augmented to
include external sources coupled to the various number currents.
Their first variation with respect to these sources give functionals
$\vew{{\cal J}_i}[\hat g]$ whose nonlocal contributions cannot be
changed by local counterterms. It is therefore sufficient to study
their gauge dependences. Here it turns out that there are gauge
dependences, but the sum of the gauge-parameter variations of the
ghost number and the Lagrange multiplier currents can again be
written as vacuum insertion of a linearized-BRS-exact operator,
which explains the findings of the explicit calculations \cite{DD,RK,KR}.

On the other hand, what Abe and Nakanishi have been doing was to
look at correlators of energy-momentum tensors. They appear to be
unconcerned with any renormalization issues and therefore are in
no position to assess the ambiguities they believe to have found.
Their alternative ``proof'' of gauge independence of the conformal anomaly
is therefore rather a collection of technical observations, which
even as such appear to be extremely limited.

There are certainly alternative approaches to studying world-sheet
anomalies, we have merely chosen one according to our preferences.
For instance, in Ref.~\cite{TT} the conformal anomaly has more recently been
determined for a non-linear de Donder gauge,\footnote{It is called
harmonic gauge in Ref.~\cite{TT}, but in the terminology of our Ref.~\cite{KR}
we have reserved this term for the background-covariant version.}
by considering the anomaly of the (full) BRS symmetry that occurs
if one insists on conformal invariance. As is well known, BRS symmetry
can be restored by local counterterms at the expense of conformal invariance.
This standard procedure has also been criticized by Abe and
Nakanishi in Ref.~\cite{AN2}, who apparently find it impossible to accept that
BRS symmetry could ever be violated. Again, they seem to disregard
the connection between the existence of anomalies and
issues of renormalization.

To summarize, we find the various criticisms put forward in Refs.~\cite{AN1,AN2}
completely baseless.

\end{document}